\documentclass[11pt]{article}
\usepackage[dvips]{graphicx}
\def\n#1{#1$\!\!${\rm\char24}\relax}

\begin{document}

\noindent
Lithuanian Journal of Physics, 2004, {\bf 44}, No. 3, pp. 
199-218 (2004)

\vspace{10mm}
\begin{center}{\bf
ATOMIC THEORY METHODS FOR THE POLARIZATION IN PHOTON AND
ELECTRON INTERACTIONS WITH ATOMS}

\vspace{ 5mm}

A. Kupliauskien\.{e}

{\small \em
Vilnius University Institute of Theoretical Physics and Astronomy, 
A,Go\v{s}tauto 12, 01108 Vilnius, Lithuania\\
E-mail: akupl@itpa.lt}

{\small Received 20 February 2004}

Dedicated to 100th anniversary of Professor A.Jucys
\end{center}

\vspace{10mm}
\begin{center}
\parbox[t]{14cm}{\small
An alternative method to the density matrix formalism for the derivation of
general expressions for the cross sections of the interaction of polarized atoms
with polarized photons and electrons is presented.
The expression for the cross section describing the polarization states of all
particles taking part in the process are obtained in the form of the expansion
via irreducible tensors that have the simplest possible behavior under
changes of directions.
The ways of the application of the general expressions suitable for the specific
experimental conditions are outlined by deriving asymmetry parameters of the
angular distributions of photoelectrons and Auger electrons following
photoionization as well as the parameters of the angular correlations between
photo- and Auger electrons.


}
\end{center}

\vspace*{ 5mm}
\noindent
{\bf 1. Introduction}
\vspace*{3mm}

In collision processes, the differential cross-section is simply a scalar with 
respect to the joint rotation of incoming and outgoing particles if neither the
projection quantum numbers of the incident particle nor the target are resolved.
For the derivation of expressions for the spectroscopic characteristics invariant
under space rotation, the powerful methods of atomic theory and angular momentum
were developed by A.Jucys and his coworkers \cite{Jucys1965,Jucys1972} in
the case of the states of atoms described with complex configurations.
These methods can be also applied for the study of
the interaction of atoms with photons, electrons and other charged particles
that is a powerful tool for the investigation of matter and interactions and have 
both theoretical and practical importance.
In any atomic process, but particularly in collisions, energy, momentum and angular
momentum are exchanged among the various constituents.
All three of these quantities are conserved, and, in a classical theory, all three
quantities may simultaneously have fixed values.
Quantum states, however, cannot be simultaneously eigenstates of linear and
angular momentum.
If the state of linear momentum is fixed, as it usually is in collision experiments,
then the angular momentum is not.
Angular momentum is conserved, but the information about it cannot be used directly.
 The mean value of products of the components
of angular momentum that are proportional to the parameters describing the
orientation and alignment  can be described.
Just these products completely specify atomic states \cite{Macek1999}.
The measurement of the parameters of the alignment and orientation helps us to 
learn about the interchange of angular momentum in atomic collisions.
Orientation and alignment parameters essentially characterize, respectively, the 
circulation of the atomic electron around the atomic core and the shape of the
excited electron cloud and its direction in space.
Orientation and alignment parameters thus allow us to go beyond the cross-section concept, 
in favorable cases leading to so called perfect scattering 
experiment, in which the sets of quantum-mechanical scattering amplitudes
and phases are completely determined \cite{Heinzmann}.

The studies of the polarization phenomena in the interactions of atoms
with charged particles and radiation stimulated the creation of  new methods
of the investigation of plasmas, ionized gaseous \cite{Kazantsev1998} and solids
\cite{Urnov1998}.
One of them is plasma polarization spectroscopy  \cite{Kazantsev1998},
since the majority of laboratory and astrophysical plasmas the electron-ion
interaction is the dominate mechanism of the radiation emission.
Measurements of the polarization
of the spectroscopic characteristics of plasma provide a unique
possibility for the diagnostics of electron and ion distribution function with high
accuracy. 
The distortion of the Maxwellian distribution function  or presence in plasma
beams may play a substantial role  for the formation of the emission spectra 
\cite{Urnov1998}.
The deviations from the Maxwellian distribution of electrons were directly
registered in laser produced \cite{Kieffer}, tokamak \cite{Fujimoto},
vacuum spark \cite{Veretenikov} and astrophysical (Solar corona) 
\cite{Kazantsev1983} plasmas by observing the polarization of line spectra
and continuous radiation.
Polarization state of emission and absorption lines could be considered as
the consequence of the polarization of atoms due to 
non equilibrium populations of magnetic sublevels or to ordering of angular 
momenta of atomic particles in plasma, that is called a self-alignment phenomena
caused by the anisotropy properties of plasma sources.

Disentanglement of the geometrical aspects of anisotropy  from the dynamical ones
is another aspect of alignment and orientation.
From the classical radiation theory, it follows that when light is observed,
only the components of the source projected onto a plane perpendicular to the 
direction of observation are imaged. 
The source must be looked at from different angles to determine its complete
electromagnetic configuration.
This holds for any measurement of radiation.
To see the complete source one must look at it from several angles.
Irreducible tensors are selected for the description of polarization
 precisely because they have the simplest possible 
behavior under changes of direction (rotation) \cite{Macek1999}.

The experiments on free polarized atoms open up the possibility to disentangle
the atomic and solid state effects \cite{Borne1997}.
Because of some atoms  \cite{Borne1997,Prumper2001}
basic practical importance for the magnetic properties
of multilayer systems and of ultra thin films on ferromagnetic substrates they are 
currently the focus of many investigations.
The photoelectron spectroscopy, exploring linear and circular magnetic
dichroism \cite{Cherepkov1995} yields detailed and site-specific information
\cite{Borne1997,Prumper2001}.
Magneto-optical effects in the VUV and soft x-ray range are very important
tool for the investigation of magnetic materials \cite{Cherepkov1995}.

To derive expressions for the parameters suitable to characterize
the polarization state in the photoionization of atoms, the density
matrix
formalism proposed by Fano and Macek \cite{Fano1957,Fano1973}
 has been widely used [16--23].
In this formalism the polarization state of  atoms or ions was described
by the statistical tensors (state multipoles) which were the basis for
the expansion of atomic density matrix \cite{Blum}.
The methods of the density matrix became common, but the work by
Fano and Macek \cite{Fano1973}  formulated
a transparent method for the light emission in the decay of a stationary state
of an atom following the excitation by collision with an electron or photon
that used a complete set of mean values of measurable
quantities instead of density matrix
to characterize a state, thereby by passing the language of the
density matrix \cite{Macek1999}.
The Wigner-Eckart theorem \cite{Jucys1965} was used to relate
measured quantities to mean values of irreducible tensors constructed
from angular momentum operators.
In effect, these mean values were proportional to state multipoles, therefore,
density matrix elements need never appear.
Then the matrix element is expressed as a product of the reduced matrix
element invariant under space rotation and the Clebsch-Gordon coefficient
depending on the orientation in space.

A further development of the ideas of Fano and Macek \cite{Fano1973}
was accomplished by Kupliauskien\.{e} {\em et al} \cite{krt2000,krt2001}.
They have used the method based on the atomic theory 
\cite{Jucys1965,Jucys1972,Rudzikas} to derive the expressions for the
photoionization cross-sections of polarized atoms by polarized radiation.
The methods of the theory of an atom \cite{Jucys1965,Jucys1972,Rudzikas}
usually were used to the isolated atom 
 for the derivation of the expressions for the spectroscopic
characteristics that are invariant under the rotation of the space.
These characteristics were made independent on the magnetic components
of the total angular momentum by using the Wigner-Eckart theorem \cite{Jucys1965}.
Then, the matrix element was expressed as a product of  reduced matrix element
that was invariant under space rotation and the Clebsch-Gordan coefficient 
depending on the orientation in space.
But the Clebsch-Gordan coefficients in the expression for the
probability or cross-section can be also used
to construct the sums of the spherical tensors for the description of the orientation
in space and rotation properties.
Until the Vilnius theoreticians have started the application of the atomic theory
methods for the investigation of the polarization in atoms \cite{krt2000,krt2001}
there have been few applications of this method,  and these have been restricted to
the polarization parameters for special cases.
Fano and Dill  \cite{Fano1972} have obtained the expression for the photoelectron
angular distribution parameter $\beta$ expressed as a sum of incoherent
contributions corresponding to the different magnitudes of the
angular momentum transferred to an unpolarized target.
The same method was extended by Klar \cite{Klar1980} to the angular distribution
of spin-polarized photoelectrons from unpolarized atoms.
The expression for the angular distribution of Auger electrons following
ionization by a beam of unpolarized electrons or protons was also
derived without the help of density matrix by Cleff and Mehlhorn
\cite{Cleff1974}.
Therefore, the formulation of the polarization theory of atoms in general
form based on
the traditional technique of spherical tensors and wave functions is
useful for people working in atomic theory.

The aim of the method based on the atomic theory approach
was the derivation of the differential cross-section and parameters 
describing the interaction 
 of polarized atoms with polarized radiation, electrons or other charged
particles in the form of the multiple expansions over spherical tensors
by using the graphical technique [30--35]
 for the integration over the angular and spin variables of
the matrix elements of the transition operators.
Any calculation made using the graphical technique can also be made using
conventional algebraic technique \cite{Brink}.
To every graphical reduction there is a corresponding algebraic reduction
because of the correspondence between graphs and algebraic formulas.
However, the graphical method has two advantages over the algebraic methods:
(a) the notation is more compact because the undetermined magnetic quantum
numbers need not be written explicitly, and (b) reductions can be made
by recognizing geometrical patterns.
The graphical technique of angular momentum
was proposed by Levinson \cite{Levinson}
to obtain an expression for the reduced matrix element invariant under the
rotation of space.
The dependence of the matrix element on the magnetic quantum numbers was
separated with the help of Wigner-Eckart theorem.
In \cite{krt2000,krt2001}, the graphical technique was extended to make it suitable
for investigating the probability of the process depending on the mutual
orientation of the particles participating in the process
or on their orientation with respect to the chosen quantization axis.

The present work is devoted to the review of the applications of the alternative
method to the density matrix formalism for the investigation of the excitation and
ionization of polarized atoms and ions by polarized radiation, electrons and other
charged  particles with subsequent decay of the formed ions and atoms in two-step
approximation.
The general expressions for the cross-sections of the excitation of atoms by
electrons as well as the radiative and dielectronic recombinations are derived
and presented for the first time.
The expression for the photoionization cross-section is also written in a more
general form more convenient for the practical applications.
The photon-atom interaction will be described in Section 2 where the excitation 
and ionization
of polarized atoms by polarized photons as well as the modifications of the
expressions for the probabilities or cross-sections in one- and multi-step
approximation will be discussed.
In Section 3, the electron-atom interactions (excitation and ionization of atoms and
ions and radiative recombination) are considered.
The radiative and Auger decay processes are described in Section 4.
The two-step process -- dielectronic recombination -- is investigated in 
Section 5.
Section 6 is devoted to demonstrate some applications of the general expressions
for the specific experimental conditions.
The review ends with concluding remarks that summarize the results and discuss
the possibilities of the application of ordinary atomic theory for the description
of the polarization in photon-atom and electron-atom interactions.

\vspace*{10mm}
\noindent
{\bf 2. Photon-atom interactions}

\vspace{ 5mm}
\noindent
{\bf 2.1. Photoexcitation of atoms}
\vspace{3mm}

The laser \cite{Golecki1999} and tunable synchrotron radiation
allows one not only
to ionize an electron of a specific outer or inner shell of an
atom but also to excite it to a specific orbital
\cite{Aksela1995,Ueda1999}.
The state of the produced atom is polarized if the radiation is 
polarized.
Thus, the excitation of atoms by polarized radiation is one of the
ways to produce atoms in polarized states for further measurements
\cite{Langer1997,Keeffe2003}.
 The photo excitation can be used as a first step to create
a resonant state with well defined angular momentum and parity for
further investigations of a polarization and angular correlation
phenomena \cite{Kab1999,Bal2000}.

 The general expression for the excitation of polarized atoms
by a polarized radiation was obtained  \cite{Kup2003} for the following process:

\begin{equation}
A(\alpha_0J_0M_0) +h\nu(\hat{\epsilon}_\lambda{\rm\bf k}_0)
\to A^*(\alpha_1J_1M_1).
\end{equation}

\noindent
Here an atom $A$ in the state $\alpha_0J_0M_0$ is excited by the
electromagnetic radiation into the state $\alpha_1J_1M_1$,
$\alpha_0$ and $\alpha_1$ define the configuration and other quantum numbers,
$J$ is the total angular momentum and $M$ is its projection.
The electromagnetic radiation is described with the wave vector {\bf k}$_0$
($k_0=|{\rm\bf k}_0|=\omega/c$, $\omega=2\pi\nu$ and $\nu$ is 
the frequency of light) and
 unit vector $\hat{\epsilon}_\lambda$ of the polarization
($\lambda$ is the helicity, $\lambda=\pm 1$).
The system of atomic units is used in the present work ($\hbar=e=m=1,
c=137$ unless these constants are displayed explicitly).
The assumption is taken into account that the fine structure splitting
$\gg$ line width $\gg$ hyperfine structure splitting.
Then the states of an atom can be specified by the total angular
momentum $J$ of all electronic shells.
The modifications enabling the calculations of the probability in the case
when hyperfine structure is important will be described below.
In the present work, it is assumed that the directions for the measurement 
of the projections  $M_0$ and $M_1$ may be different.

The differential cross section of the process (1) can be written  as follows
\cite{Bal2000}:
$$
\frac{d\sigma(\alpha_0J_0M_0\hat{\epsilon}_\lambda{\rm\bf k}_0 \to
 \alpha_1J_1M_1)}{ d\Omega}
= 2\pi^2
\left[ \int \langle \alpha_1J_1M_1|{\rm \bf  J}({\rm \bf  r})|
\alpha_0J_0M_0\rangle
{\rm\bf A}_{\lambda {\rm\bf  k}_0}({\rm \bf  r}) d{\rm \bf  r}\right]
$$
\begin{equation}
\times
\left[ \int \langle \alpha_1J_1M_1|{\rm \bf  J}({\rm \bf  r})|
\alpha_0J_0M_0\rangle
{\rm\bf A}_{\lambda {\rm\bf k}_0}({\rm \bf  r}) d{\rm \bf  r}\right]^*.
\end{equation}

\noindent
Here  ${\rm \bf  J}({\rm \bf  r})$ stands for the operator of the current of 
electrons and ${\rm\bf A}_{\lambda {\rm\bf k}_0}({\rm \bf  r})$ is the operator of 
the vector potential of  electromagnetic field.
Taking into account that $kr \ll 1$ and the case of
 an arbitrary direction of the incidence {\bf k}$_0$ of the photon, 
and inserting the multipole expansion for 
the ${\rm\bf A}_{\lambda {\rm\bf k}_0}({\rm \bf  r})$, the expression in the right brackets
of Eq.~(2) acquires the form \cite{Bal2000}:
$$
 \int \langle \alpha_1J_1M_1|{\rm \bf  J}({\rm \bf  r})|\alpha_0J_0M_0\rangle
{\rm\bf  A}_{\lambda k_0}({\rm \bf  r}) d{\rm \bf  r}
= \sum_{p=0,1}\sum_{k=1}^{\infty}\sum_{q=-k}^{q=k}
i^k(-i\lambda)^p \left[\frac{k+1}{k}\right]^{1/2}
\frac{k_0^{k-1/2}}{(2k-1)!!} D^k_{q\lambda}(\hat{k}_0)
$$
\begin{equation}
\times
\langle \alpha_1 J_1M_1|{\cal Q}^p_{kq}|\alpha_0 J_0M_0\rangle 
=\sum_{k=1}^{\infty}\sum_{q=-k}^{q=k}
\langle \alpha_1 J_1M_1|Q^{(k)}_q|\alpha_0 J_0M_0\rangle 
 D^k_{q\lambda}(\hat{k}_0),
\end{equation}
\begin{equation}
\langle \alpha_1 J_1M_1|Q^{(k)}_q|\alpha_0 J_0M_0\rangle
=  k_0^{k-1/2} \sum_{p=0,1} \left[\frac{k+1}{k}\right]^{1/2}
\frac{i^k(-i\lambda)^p }{(2k-1)!!}
\langle \alpha_1 J_1M_1|{\cal Q}^p_{kq}|\alpha_0 J_0M_0\rangle.
\end{equation}

\noindent
Here $ D^k_{q\lambda}(\hat{k_0})$ is the Wigner rotation matrix \cite{VMK1988}
for transforming from the helicity frame ($\hat{k}_0$ is
the direction of incoming radiation) to the frame common to all
particles participating in the process and used for the evaluation of 
reduced matrix elements.
$p=0$ indicates the operator of the electric multipole transition (E$k$) 
  \cite{Rudzikas}
\begin{equation}
{\cal Q}^0_{kq}= - r^k C^{(k)}_q
\end{equation}
and, $p=1$ shows the operator of the magnetic multipole transition (M$k$)
\begin{equation}
{\cal Q}^1_{k q}
= -\frac{1}{c} [k(2k-1)]^{1/2}r^{k-1}
\left\{ \frac{1}{k+1}\left[ C^{(k-1)} \times  L^{(1)}
\right]^{(k)}_q
+\left[C^{(k-1)} \times S^{(1)}\right]^{(k)}_q \right\}.
\end{equation}

\noindent
Here  $L^{(1)}$ and $S^{(1)}$ are the operators of the orbital and
spin  angular momentum, respectively,
$C^{(k)}_q$ is the operator of the spherical function normalized to
$[4\pi/(2k+1)]^{1/2}$ \cite{Jucys1965}.

Note that the parities of the magnetic and electric multipole fields
are $(-1)^k$ and $(-1)^{k+1}$, respectively. Only the magnetic
 (M$k$) and electric  (E$k$) part contributes between specific
electronic states owing to parity selection rules. Since we are 
considering pure photon states, there is no need to introduce
Stokes parameters explicitly.
In electrical dipole approximation, the matrix element (4) is as 
follows
\begin{equation}
\langle \alpha_1 J_1M_1|Q^{(1)}_q|\alpha_0 J_0M_0\rangle
=\sqrt{2 k_0}
\langle \alpha_1 J_1M_1|{\cal Q}^0_{1q}|\alpha_0 J_0M_0\rangle.
\end{equation}

The helicity $\lambda=\pm 1$ describes the right-hand and left-hand
circular  radiation. 
In the case of any polarization $\epsilon$, the polarization of radiation may be
expressed via the circular polarization. Then
\begin{equation}
{\rm\bf  A}_{\epsilon {\rm\bf k}_0}({\rm \bf  r})=
\alpha {\rm\bf  A}_{\lambda=+1 {\rm\bf k}_0}({\rm \bf  r})
+\beta {\rm\bf  A}_{\lambda=-1 {\rm\bf k}_0}({\rm \bf  r})
\end{equation}

\noindent
that has to be inserted into Eq.~(3) to obtain the expression 
for the excitation cross section (2).

Sometimes it is more convenient to analyze the polarization state
of the particle with respect to the direction that differs from the one used 
for the calculation of the matrix element where all particles
of the process (1) should be described in the same coordinate system
and their projections of the angular momentum on the same
quantization axis.
For the transfer from the wave function $|JM\rangle$ defined in the
laboratory fixed direction to the wave function $|J\tilde{M}\rangle$
of the atomic frame used for the evaluation of reduced matrix elements, 
the coordinate rotation transformation
\cite{VMK1988} is used
\begin{equation}
|JM\rangle=\sum_{\tilde{M}}D^J_{\tilde{M}M}(\hat{J})\;
|J\tilde{M}\rangle.
\end{equation}

\noindent
Here $D^J_{\tilde{M}M}(\hat{J})$ denotes the Wigner rotation
matrix \cite{VMK1988}, and the hat on $J$ indicates the rotation by
solid angle that transforms the atomic frame into the laboratory one
used for the polarization measurements of the characteristics depending 
on {\rm J}.
In the laboratory system of coordinates, the matrix element
$\langle \alpha_1J_1M_1|Q^{(k)}_q|\alpha_0J_0M_0\rangle$ can be
written by taking into account Eq.~(9)  in the form:
\begin{equation}
\langle \alpha_1J_1M_1|Q^{(k)}_q|\alpha_0J_0M_0\rangle =
\sum_{\tilde{M}_0,\tilde{q},\tilde{M}_1}
\langle \alpha_1J_1\tilde{M}_1|Q^{(k)}_{\tilde{q}}|
\alpha_0J_0\tilde{M}_0\rangle
D^{J_0}_{\tilde{M}_0 M_0}(\hat{J}_1)\;
D^{k}_{\tilde{q}q}(\hat{k}_0)\;
D^{*J_1}_{\tilde{M}_1 M_1}(\hat{J}_1).
\end{equation}

\noindent
The matrix element
$\langle \alpha_1J_1\tilde{M}_1|Q^{(k)}_{\tilde{q}}|
\alpha_0J_0\tilde{M}_0\rangle$ is defined in the atomic frame.
The angular momentum part of this matrix element was obtained in \cite{Kup2003}
with the help of the graphical technique of the angular momentum 
\cite{Jucys1965}.
The final expression for the excitation cross-sections is:

$$
\frac{d\sigma (\alpha_0J_0M_0\hat{\epsilon}_q{\rm\bf k}_0
\to \alpha_1J_1M_1)}{d\Omega} =
$$
$$
C\sum_{K_0,K_r,K_1,k,k'}
\frac{1}{2K_1+1} {\cal A}^r(K_0,K_r,K_1,k,k')
\sum_{N_0,N_r,N_1,N'_1}
\left[\begin{array}{ccc}
K_0&K_r&K_1\\N_0&N_r&N_1
\end{array}\right]
$$
\begin{equation}
\times
T^{*K_0}_{N_0}(J_0,J_0,M_0|\hat{J}_0)\;
T^{*K_r}_{N_r}(k,k',q|\hat{k}_0)\;
T^{K_1}_{N_1 N'_1}(J_1,J_1,M_1,M_1|\hat{J}_1) 
\end{equation}

\noindent
where
$$
{\cal A}^r(K_0,K_r,K_1,k,k')=
(\alpha_1J_1||Q^{(k)}||\alpha_0J_0)(\alpha_1J_1||Q^{(k')}||\alpha_0J_0)^*
\left\{\begin{array}{ccc}
J_0&K_0&J_0\\k&K_r&k'\\J_1&K_1&J_1
\end{array}\right\}
$$
\begin{equation}
[(2J_0+1)(2J_1+1)(2k+1)(2K_1+1)]^{1/2},
\end{equation}
\begin{equation}
T^{*K}_{N}(J,J',M|\hat{J})=
(-1)^{J'-M}\left[\frac{4\pi}{2J+1}\right]^{1/2}
\left[\begin{array}{ccc}
J&J'&K\\M&-M&0
\end{array}\right]
Y^*_{KN}(\theta,\phi),
\end{equation}
\begin{equation}
T^{K_1}_{N_1N'_1}(J_1,J'_1,M_1,M'_1|\hat{J}_1)=(-1)^{J'_1-M'_1}
\left[\frac{2K_1+1}{2J_1+1}\right]^{1/2}
\left[\begin{array}{ccc}
J_1 & J'_1 & K_1\\ M_1 & M'_1 & N'_1
\end{array} \right]
D^{*K_1}_{N_1N'_1}(\hat{J}_1) .
\end{equation}

\noindent
In Eq.~(12), the relation
\begin{equation}
(\alpha_1J_1||Q^{(k)}||\alpha_0J_0)=[2J_1+1]^{1/2}
\langle\alpha_1J_1||Q^{(k)}||\alpha_0J_0\rangle
\end{equation}

\noindent
is taken into account. In Eq.~(11), $C=2\pi^2$.
$\hat{J}$ denotes the angles of {\bf J} with respect to the $z$ axis
of laboratory frame.
In the case when hyperfine structure is important, the reduced
matrix element $(\alpha_1J_1||Q^{(k)}||\alpha_0J_0)$
should be changed by $(\alpha_1J_1(I)F_1||Q^{(k)}||\alpha_0J_0(I)F_0)$
in Eq.~(12).
A simple relation between these two matrix elements holds:
$$
(\alpha_1J_1(I)F_1||Q^{(k)}||\alpha_0J_0(I)F_0)=
(-1)^{F_0-J_1+I+k}[(2F_0+1)(2J_1+1)]^{1/2}
$$
\begin{equation}
\times
\left\{\begin{array}{ccc}
F_0&k&F_1\\J_1&I&J_0 \end{array}\right\}
(\alpha_1J_1||Q^{(k)}||\alpha_0J_0).
\end{equation}

\noindent
$I$ is the spin of the nucleus.
The values of $J_0,J_1$ in (11)--(15) should be changed by $F_0,F_1$.
The probability $W (\alpha_0J_0M_0\hat{\epsilon}_q{\rm\bf k}_0
\to \alpha_1J_1M_1)$
equals to the cross section (2) divided by the density of the flow of the
radiation. Thus the same expression of the cross section can be used
for the probability by changing  only the definition of the constant $C$.

\vspace*{ 5mm}
\noindent
{\bf 2.2. Photoionization of atoms}
\vspace{3mm}

The expression for the differential cross-section of the photoionization process
\begin{equation}
A(\alpha_0J_0M_0) + h\nu (\epsilon_{q1}{\rm \bf k}_0,{\hat \epsilon}) \to
A^+(\alpha_1J_1M_1) + {\rm e^-}({\rm\bf p}_1,sm_s)
\end{equation}

\noindent
was derived by Kupliauskien\.{e} {\em et al} \cite{krt2001} in the
case of dipole approximation and the levels $LS(J)IF$ of an atom.
Here the ejected photoelectron has momentum {\bf p}$_1$, and the projection of its
spin $s$ is indicated by $m_s$.

In the case when the hyperfine structure is small and taking into account of all 
multipoles,
the general expression for the differential cross-section of Eq.~(17) can be written
as follows:
$$
\frac{d\sigma(J_0M_0 \hat{\epsilon}_{q1} {\rm\bf k}_{01}
\to J_1M_1{\rm\bf p}_1m_s)}{d\Omega_{p_1}}
$$
$$
=
\pi
\sum_{K_0,K_r,K_1,K_\lambda,K_s,K_j,K,k_1,k'_1}
{\cal B}^{ph}(K_1,K_0,K_r,K_\lambda,K_s,K_j,K,k_1,k'_1)
$$
$$
\times
\sum_{N_0,N_r,N_1,N_\lambda,N_j,N,N_s}
          \left[
\begin{array}{ccc}
K_1 & K_j & K \\ N_1 &N_j & N
\end{array}
          \right]
          \left[
\begin{array}{ccc}
K_\lambda & K_s & K_j \\ N_\lambda &N_s & N_j
\end{array}
           \right]
          \left[
\begin{array}{ccc}
K_0 & K_r & K \\ N_0 &N_r & N
\end{array}
           \right]
T^{K_1}_{N_1 N'_1}(J_1,J_1,M_1,M_1|\hat{J}_1)
$$
\begin{equation}
\times
T^{*K_0}_{N_0}(J_0,J_0,M_0|\hat{J}_0) \;
T^{*K_r}_{N_r}(k_1,k'_1,q_1|\hat{{\rm\bf k}}_{01}) \;
T^{K_s}_{N_s}(s,s,m_s|\hat{s}) \;
\sqrt{4\pi}\;Y_{K_\lambda N_\lambda}(\hat{{\rm \bf p}}_1),
\end{equation}

$$
{\cal B}^{ph}(K_1,K_0,K_r,K_\lambda,K_s,K_j,K,k_1,k'_1)=
 \sum_{\lambda,j,J,\lambda',j',J'}(2J+1)(2J'+1)
(-1)^{\lambda'}
$$
$$
\times
\langle \alpha_1 J_1\varepsilon_1\lambda(j)J||Q^{(k_1)}||\alpha_0
J_0\rangle
\langle \alpha_1 J_1\varepsilon_1\lambda'(j')J'||Q^{(k'_1)}||
\alpha_0J_0\rangle^*
$$
$$
\times
[(2J_0+1)(2K_j+1)((2J_1+1)(2k_1+1)(2s+1)(2\lambda+1)
(2\lambda'+1)(2j+1)(2j'+1)]^{1/2}
$$
\begin{equation}
\times
         \left[
\begin{array}{ccc}
\lambda &\lambda' & K_\lambda\\ 0 & 0 & 0
\end{array}
          \right]
         \left\{
\begin{array}{ccc}
J_0 &K_0 & J_0\\ k_1 & K_r & k'_1\\ J' & K & J
\end{array}
          \right\}
         \left\{
\begin{array}{ccc}
J_1 &K_1 & J_1\\ j' & K_j & j  \\ J' & K & J
\end{array}
          \right\}
         \left\{
\begin{array}{ccc}
\lambda' &K_\lambda & \lambda\\ s & K_s & s  \\ j' & K_j & j
\end{array}
          \right\}.
\end{equation}

In the case of the photoionization as  the first step process, the expression for the
cross-section should be modified (see Section 2.3). 
It acquires the form:

$$
\frac{d\sigma_{K_1N_1}(\alpha_0J_0M_0 \hat{\epsilon}_{q1}{\rm\bf k}_{01}
\to \alpha_1J_1{\rm\bf p}_1m_s)}
{d\Omega_{p_1}}
$$
$$
= 
\pi
\sum_{K_0,K_r,K_\lambda,K_s,K_j,K,k_1,k'_1}
{\cal B}^{ph}(K_1,K_0,K_r,K_\lambda,K_s,K_j,K,k_1,k'_1)
$$
$$
\times
\left[ \frac{2K_1+1}{2J_1+1}\right]^{1/2}
\sum_{N_0,N_r,N_\lambda,N_j,N,N_s}
          \left[
\begin{array}{ccc}
K_1 & K_j & K \\ N_1 &N_j & N
\end{array}
          \right]
          \left[
\begin{array}{ccc}
K_\lambda & K_s & K_j \\ N_\lambda &N_s & N_j
\end{array}
           \right]
          \left[
\begin{array}{ccc}
K_0 & K_r & K \\ N_0 &N_r & N
\end{array}
           \right]
$$
\begin{equation}
\times
T^{*K_0}_{N_0}(J_0,J_0,M_0|\hat{J}_0) \;
T^{*K_r}_{N_r}(k_1,k'_1,q_1|\hat{{\rm\bf k}}_{01}) \;
T^{K_s}_{N_s}(s,s,m_s|\hat{s}) \;
\sqrt{4\pi}\;Y_{K_\lambda N_\lambda}(\hat{{\rm \bf p}}_1).
\end{equation}

Further the general expressions (18)--(20) can be used to obtain some
special expressions for specific experimental conditions
also investigated by other authors with the help of the density matrix formalism,
e.g., the expressions for the
angular  distribution and spin polarization of photoelectrons
 in the case of nonpolarized atom.

\vspace*{ 5mm}
\noindent
{\bf 2.3. One- and multi-step processes}
\vspace{3mm}

The excitation or ionization  of an atom by laser or other electromagnetic
radiation often is used to prepare it in a polarized state for further investigation.
Then the magnetic state of an atom or ion in the final state $J_1 M_1$
 is not observed, and summation of $M_1$ has to be performed coherently.
In this case, the excitation (1) is the first step of the multi-step
process while the second step is
\begin{equation}
A^*(\alpha_1J_1M_1)+b(\alpha)\to A(\alpha_2J_2M_2) +
b'(\alpha').
\end{equation}

\noindent
Here $b(\alpha)$ stands for impacting particle or
electromagnetic radiation in the state $\alpha$ and $b'(\alpha')$
indicates ionized and emitted one or more particles.
In two-step approximation, the probability of both processes (1) and
(21) can be written   as coherent sum since the projection
$M_1$ cannot be observed \cite{Klar1980}:

$$
W (\alpha_0J_0M_0\hat{\epsilon}{\rm\bf  k}_0 \to \alpha_1J_1 \alpha \to
\alpha_2J_2M_2\alpha')= C' |\sum_{M_1}
\langle \alpha_2J_2M_2\alpha'|H_2|\alpha_1J_1M_1\alpha\rangle
\langle \alpha_1J_1M_1|H_1|\alpha_0J_0M_0\rangle|^2
$$
$$
=C'
\sum_{M_1,M'_1}
\langle \alpha_2J_2M_2\alpha'|H_2|\alpha_1J_1M_1\alpha\rangle
\langle \alpha_2J_2M_2\alpha'|H_2|\alpha_1J_1M'_1\alpha\rangle^*
$$
\begin{equation}
\times
\langle \alpha_1J_1M_1|H_1|\alpha_0J_0M_0\rangle
\langle \alpha_1J_1M'_1|H_1|\alpha_0J_0M_0\rangle^*.
\end{equation}

\noindent
The matrix elements in Eq.~(22) are defined similar to 
Eq.~(10) allowing the measurement
in the direction different from that used for evaluation.
$H_1$ and $H_2$ are the operators of the interaction in the first and second
processes, respectively. 
Then the product  of two Wigner rotation matrices coming up from the matrix
element and its complex conjugate  in Eq.~(22)  should be replaced by
\begin{equation}
D^{J_1}_{\tilde{M}_1 M_1}(\hat{J}_1) D^{*J'_1}_{\tilde{M}'_1 M'_1}(\hat{J}_1)
=\sum_{K_1,N_1,N'_1}
\left[\begin{array}{ccc}
J'_1 & K_1 & J_1 \\ \tilde{M}'_1 & N_1 & \tilde{M}_1
\end{array} \right]
T^{*K_1}_{N_1N'_1}(J_1,J'_1,M_1,M'_1|\hat{J}_1),
\end{equation}

\noindent
and the summation over $M_1$ and $M'_1$ in Eq.~(22) is possible to be carried out.
 From the
examination of the expressions (29) in \cite{kt2003a} and (11) and (18) of the present 
work for both terms in Eq.~(22) it follows that only the tensors
$T^{K_1}_{N_1N'_1}(J_1,J'_1,M_1,M'_1|\hat{J}_1)$ depend 
on $M_1$ and $M'_1$.
The sum over $M_1,M'_1$ of the product  of these tensors is equal to
$$
\sum_{M_1,M'_1} T^{K_1}_{N_1N}(J_1,J_1,M_1,M'_1|\hat{J_1}) \;
T^{*K'_1}_{N'_1N'}(J_1,J_1,M_1,M'_1|\hat{J_1})
$$
$$
=
\frac{\sqrt{(2K_1+1)(2K'_1+1)}}{2J_1+1}
D^{*K_1}_{N_1N}(\hat{J}) D^{K'_1}_{N'_1N'}(\hat{J})
\sum_{M_1,M'_1}
(-1)^{2J_1-2M'_1}
\left[ \begin{array}{ccc}
J_1 & J_1 & K_1\\ M_1 & M'_1 & N
\end{array} \right]
\left[ \begin{array}{ccc}
J_1 & J_1 & K'_1\\ M_1 & M'_1 & N'
\end{array} \right]
$$
\begin{equation}
=\frac{2K_1+1}{2J_1+1}
D^{*K_1}_{N_1N}(\hat{J}) D^{K_1}_{N'_1N}(\hat{J})
\delta(K_1,K'_1) \delta(N,N').
\end{equation}

\noindent
Then  the quantization axis can be chosen along  the z axis of the laboratory
coordinate system, and
$D^{*K_1}_{N_1N}(0,0,0) D^{K_1}_{N'_1N}(0,0,0)=
\delta(N_1,N)\delta(N'_1,N)$
The square root of the multiple $(2K_1+1)/(2J_1+1)$
is convenient to  attribute to both terms of the following expression:
$$
W (\alpha_0J_0M_0\hat{\epsilon}{\rm\bf k}_0
\to \alpha_1J_1\alpha  \to
\alpha_2J_2M_2\alpha')
$$
\begin{equation}
=
\sum_{K_1,N_1}
W_{K_1N_1}(\alpha_0J_0M_0\hat{\epsilon}{\rm\bf k}_0
\to \alpha_1J_1\alpha)
\cdot
W^A_{K_1N_1}(\alpha_1J_1\alpha \to
\alpha_2 J_2M_2\alpha') .
\end{equation}

\noindent
Here the sum over $M_1,M'_1$ in Eq.~(22) is changed by the sum over $K_1,N_1$, 
i.e. the probability of the two-step process is
expanded as the sum of state multipoles.
For example, the expressions for the photoexcitation probability (11) and 
that of the second process slightly change.
For the excitation probability, it is
$$
 W_{K_1N_1}(\alpha_0J_0M_0\hat{\epsilon}_q{\rm\bf k}_0
\to \alpha_1J_1)= 
\frac{C}{[2J_1+1]^{1/2}} 
 \sum_{K_0,K_r,k,k'}  B^r(K_0,K_r,K_1,k,k')
\sum_{N_0,N_r,q}
\left[\begin{array}{ccc}K_0&K_r&K_1\\N_0&N_r&N_1
\end{array}\right]
$$
\begin{equation}
\times
T^{*K_0}_{N_0}(J_0,J_0,M_0|\hat{J}_0)
T^{*K_r}_{N_r}(k,k',q|\hat{k}_0),
\end{equation}
\begin{equation}
W_{00}(\alpha_0J_0k\to \alpha_1J_1)= 
\frac{C}{(2J_0+1)(2k+1)[2J_1+1]^{1/2}}
|\langle  \alpha_1J_1||Q^{(k)}||\alpha_0J_0\rangle|^2.
\end{equation}

\noindent
The expression for the second term in Eq.~(25) depends on the second-step
process. In the case of the Auger decay, it is presented by 
Kupliauskien\.{e} and Tutlys (see Eq. (4) in \cite{KT2003}).

The proposed method is easy to generalize for multistep process when 
intermediate states are not observed. In the case of tree-step process
where the fluorescence radiation of the doubly charged ion formed following
the photoionization of an atom and Auger decay of the singly charged
ion is registered, the summation over intermediate states gives
$$
\sum_{M_1,M'_1,M_2,M'_2}
W^1(J_0M_0{\rm\bf k}_1\to J_1M_1M'_1{\rm\bf p}_1m_1)
W^2(J_1M_1M'_1{\rm\bf p}_1m_1\to J_2M_2M'_2{\rm\bf p}_2m_2)
$$
$$
\times
W^3(J_2M_2M'_2{\rm\bf p}_2m_2 \to J_3M_3{\rm\bf k}_2)
$$
$$
=\sum_{K_1,N_1,K_2,N_2}
W^1_{K_1N_1}(J_0M_0{\rm\bf k}_1\to J_1{\rm\bf p}_1m_1)
\frac{2K_1+1}{2J_1+1}
$$
\begin{equation}
\times
W^2_{K_1N_1K_2N_2}(J_1{\rm\bf p}_1m_1\to J_2{\rm\bf p}_2m_2)
\frac{2K_2+1}{2J_2+1}
W^3_{K_2N_2}(J_2{\rm\bf p}_2m_2 \to J_3M_3{\rm\bf k}_2).
\end{equation}
\noindent
The square root of each multiple $(2K+1)/(2J+1)$ in Eq.~(28)
is also convenient to  attribute to both neighbouring terms.

\vspace*{ 5mm}
\noindent
{\bf 3. Electron-atom interactions}

\vspace{ 5mm}
\noindent
{\bf 3.1. Excitation of atoms by electrons}
\vspace{3mm}

The process of the excitation of polarized atoms by polarized electrons
can be written as follows:
\begin{equation}
A(\alpha_0 J_0M_0)+e^-({\rm\bf p}_0 m_0)\to
A(\alpha_1 J_1M_1) +e^-({\rm\bf p}_1 m_1).
\end{equation}

The expression for the differential cross-section of the process (29)
 is easy to obtain by the 
method described  in  \cite{krt2001,kt2003a} and is as follows:
$$
\frac{d^2\sigma(\alpha_0J_0M_0{\rm \bf p}_0 m_0 \to
\alpha_1J_1M_1 {\rm \bf p}_1m_1)}
{ d\varepsilon_2 d\Omega_1 }
$$
$$
= 4\pi  C
\sum_{\begin{array}{c}K,K_0,K'_0,K_{\lambda 0},K_{s0},K_1\\K'_1,
K_{\lambda 1},K_{s 1}\end{array}}
{\cal B}^{ex}(K_0,K'_0,K_1,K'_1,K_{\lambda 0},K_{s0},K_{\lambda 1},K_{s 1},
K)
$$
$$
\times
\sum_{\begin{array}{c}N_0,N'_0,N_{\lambda 0},N_{s0},N_1\\N'_1,
N_{\lambda 1},N_{s 1},N\end{array}}
\left[ \begin{array}{c c c}
K_{\lambda 0}&K_{s 0}&K'_0\\N_{\lambda 0}&N_{s 0}&N'_0
\end{array}\right]
\left[ \begin{array}{c c c}
K_0&K'_0&K\\N_0&N'_0&N
\end{array}\right]
\left[ \begin{array}{c c c}
K_1&K'&K\\N_1&N'&N
\end{array}\right]
$$
$$
\times
\left[ \begin{array}{c c c}
K_1&K'_1&K\\N_1&N'_1&N
\end{array}\right]
\left[ \begin{array}{c c c}
K_{\lambda 1}&K_{s 1}&K'_1\\N_{\lambda 1}&N_{s 1}&N'_1
\end{array}\right]
Y^*_{K_{\lambda 0}N_{\lambda 0}}(\hat{p}_0)\;
Y_{K_{\lambda 1}N_{\lambda 1}}(\hat{p}_1)\;
$$
\begin{equation}
\times
T^{*K_0}_{N_0}(J_0,J_0,M_0|\hat{J}_0)\;
T^{K_1}_{N_1}(J_1,J_1,M_1|\hat{J}_1)\;
T^{*K_{s0}}_{N_{s0}}(s,s,m_0|\hat{s})\;
T^{K_{s1}}_{N_{s1}}(s,s,m_1|\hat{s})
\end{equation}

$$
{\cal B}^{ex}(K_0,K'_0,K_1,K'_1,K_{\lambda 0},K_{s0},
K_{\lambda 1},K_{s 1},K)
$$
$$
=
\sum_{
\lambda_0,\lambda'_0,\lambda_1,\lambda'_1,j_0,j'_0,j_1,j'_1,J,J'}
(2J+1)(2J'+1)(2K+1)(2s+1)(-1)^{\lambda'_0+\lambda'_1}
$$
$$
\times
\langle\alpha_1J_1,\varepsilon_1\lambda_1(j_1)J||H||\alpha_0J_0,
\varepsilon_0\lambda_0(j_0)J\rangle
\langle\alpha_1J_1,\varepsilon_1\lambda'_1(j'_1)J'||H||\alpha_0J_0,
\varepsilon_0\lambda'_0(j'_0)J'\rangle^*
$$
$$
\times
[(2\lambda_0+1) (2\lambda'_0+1)(2\lambda_1+1) (2\lambda'_1+1)
(2j_0+1)(2j'_0+1)(2j_1+1)(2j'_1+1)
$$
$$
\times
(2J_0+1)(2J_1+1)(2K'_0+1)(2K'_1+1)]^{1/2}
\left[ \begin{array}{c c c}
\lambda_0&\lambda'_0&K_{\lambda 0}\\0&0&0
\end{array}\right]
\left[ \begin{array}{c c c}
\lambda_1&\lambda'_1&K_{\lambda 1}\\0&0&0
\end{array}\right]
$$
\begin{equation}
\times
\left\{ \begin{array}{c c c}
J_0&K_0&J_0\\j'_0&K'_0&j_0\\J'&K&J
\end{array}\right\}
\left\{ \begin{array}{c c c}
\lambda'_0&K_{\lambda 0}&\lambda_0\\s&K_{s0}&s\\j'_0&K'_0&j_0
\end{array}\right\}
\left\{ \begin{array}{c c c}
\lambda'_1&K_{\lambda 1}&\lambda_1\\s&K_{s1}&s\\j'_1&K'_1&j_1
\end{array}\right\}
\left\{ \begin{array}{c c c}
J_1&K_1&J_1\\j'_1&K'_1&j_1\\J&K&J'
\end{array}\right\} .
\end{equation}

In Eq.~(31), $\langle\alpha_1J_1,\varepsilon_1\lambda_1(j_1)J||H||\alpha_0J_0,
\varepsilon_0\lambda_0(j_0)J\rangle$ is the reduced matrix element of the 
electrostatic interaction.
If the excitation of atoms by electrons is used as the first step process for 
the preparation of atoms in polarized excited states for the next step process,
the projections $M_1$ are not registered.
Then the summation over $M_1$ should be performed following the
recommendations of Section 2.3.

\vspace{ 10mm}
\noindent
{\bf 3.2. Ionization of atoms by electrons}
\vspace{3mm}

If neither the projection quantum numbers of the incident electron
nor the target are resolved, the differential cross-section for the
ionization of atoms by electrons is simply a scalar with respect
to the joint rotation of the incoming and outgoing electron momenta
{\bf p}$_0$, {\bf p}$_1$, and {\bf p}$_2$, respectively.
This basic symmetry is destroyed by an initial orientation and/or
alignment of the target, and can supply more information about the
elementary processes in such diverse fields as discharge and plasma
physics \cite{Serapinas1994}, fusion physics \cite{Fujimoto}, and the physics
and chemistry of the upper atmosphere \cite{Dorn1998}.
For the  interpretation of recently measured ionization
cross-sections  of polarized Na atoms \cite{Dorn1998,Lower2001},
the expressions enabling one to describe the polarization states
of all particles taking part in the process are necessary.

Recently the general expression for the description of the ionization of 
polarized atoms by polarized electrons 
\begin{equation}
A(\alpha_0 J_0M_0)+e^-({\rm\bf p}_0 m_0)\to
A^+(\alpha_1 J_1M_1)+e^-({\rm\bf p}_2 m_2) +e^-({\rm\bf p}_1 m_1).
\end{equation}
was obtained by Kupliauskien\.{e} and  Glem\v{z}a \cite{kg2003}.
In Eq.~(32), {\bf p}$_i$ denotes the momentum of the electron in the initial
($i=0$) and final ($i=1,2$) states, $m_i$ indicates the projection
of an electron spin.
The fine structure splitting was assumed larger than hyperfine one.
The expression for the cross-section
 was derived by using the graphical technique of the angular momentum
 \cite{Jucys1965} and is as follows:
$$
\frac{d^3\sigma(\alpha_0J_0M_0{\rm \bf p}_0 m_0 \to
\alpha_1J_1M_1{\rm \bf p}_2 m_2 {\rm \bf p}_1m_1)}
{ d\varepsilon_2 d\Omega_1 d\Omega_2}
$$
$$
= C (4\pi)^{3/2}
\sum_{\begin{array}{c}K,K_0,K'_0,K_{\lambda 0},K_{s0},K',K_1\\K'_1,K'_2,
K_{\lambda 1},K_{s 1},K_{\lambda 2},K_{s 2}\end{array}}
{\cal B}^{ion}(K_0,K'_0,K,K_{\lambda 0},K_{s0},K_1,K',K_{\lambda 1},K_{s 1},
K'_1,K_{\lambda 2},K_{s 2},K'_2)
$$
$$
\times
\sum_{\begin{array}{c}N,N_0,N'_0,N_{\lambda 0},N_{s0},N',N_1\\N'_1,N'_2,
N_{\lambda 1},N_{s 1},N_{\lambda 2},N_{s 2}\end{array}}
\left[ \begin{array}{c c c}
K_{\lambda 0}&K_{s 0}&K'_0\\N_{\lambda 0}&N_{s 0}&N'_0
\end{array}\right]
\left[ \begin{array}{c c c}
K_0&K'_0&K\\N_0&N'_0&N
\end{array}\right]
\left[ \begin{array}{c c c}
K_1&K'&K\\N_1&N'&N
\end{array}\right]
$$
$$
\times
\left[ \begin{array}{c c c}
K'_2&K'_1&K'\\N'_2&N'_1&N'
\end{array}\right]
\left[ \begin{array}{c c c}
K_{\lambda 1}&K_{s 1}&K'_1\\N_{\lambda 1}&N_{s 1}&N'_1
\end{array}\right]
\left[ \begin{array}{c c c}
K_{\lambda 2}&K_{s 2}&K'_2\\N_{\lambda 2}&N_{s 2}&N_2
\end{array}\right]
Y^*_{K_{\lambda 0}N_{\lambda 0}}(\hat{p}_0)\;
Y_{K_{\lambda 1}N_{\lambda 1}}(\hat{p}_1)\;
$$
$$
\times
Y_{K_{\lambda 2}N_{\lambda 2}}(\hat{p}_2)\;
T^{*K_0}_{N_0}(J_0,J_0,M_0|\hat{J}_0)\;
T^{K_1}_{N_1}(J_1,J_1,M_1|\hat{J}_1)\;
T^{*K_{s0}}_{N_{s0}}(s,s,m_0|\hat{s})\;
$$
\begin{equation}
\times
T^{K_{s1}}_{N_{s1}}(s,s,m_1|\hat{s})\;
T^{K_{s2}}_{N_{s2}}(s,s,m_2|\hat{s}) ,
\end{equation}

$$
{\cal B}^{ion}(K_0,K'_0,K,K_{\lambda 0},K_{s0},K_1,K',K_{\lambda 1},K_{s 1},
K'_1,K_{\lambda 2},K_{s 2},K'_2)
$$
$$
=
\sum_{\begin{array}{c}
\lambda_0,\lambda'_0,\lambda_1,\lambda'_1,\lambda_2,\lambda'_2,
j_0,j'_0\\
j_1,j'_1,j_2,j'_2,J,J',j,j'\end{array}}
(2J+1)(2J'+1)(2K+1)(2s+1)(-1)^{\lambda'_0+\lambda'_1+\lambda'_2}
$$
$$
\times
\langle\alpha_1J_1,\varepsilon_2\lambda_2(j_2)
\varepsilon_1\lambda_1(j_1)j,J||H||\alpha_0J_0,
\varepsilon_0\lambda_0(j_0)J\rangle
$$
$$
\times
\langle\alpha_1J_1,\varepsilon_2\lambda'_2(j'_2)
\varepsilon_1\lambda'_1(j'_1)j',J'||H||\alpha_0J_0,
\varepsilon_0\lambda'_0(j'_0)J'\rangle^*
$$
$$
\times
[(2s+1)(2\lambda_0+1) (2\lambda'_0+1)(2\lambda_1+1) (2\lambda'_1+1)
(2\lambda_2+1) (2\lambda'_2+1)(2j_0+1)(2j'_0+1)
$$
$$
\times
(2j_1+1)(2j'_1+1)
(2j_2+1)(2j'_2+1)(2j+1)(2j'+1)(2J_0+1)(2J_1+1)(2K'_0+1)(2K'_1+1)
$$
$$
\times
(2K'+1)(2K'_2+1)]^{1/2}
\left[ \begin{array}{c c c}
\lambda_0&\lambda'_0&K_{\lambda 0}\\0&0&0
\end{array}\right]
\left[ \begin{array}{c c c}
\lambda_1&\lambda'_1&K_{\lambda 1}\\0&0&0
\end{array}\right]
\left[ \begin{array}{c c c}
\lambda_2&\lambda'_2&K_{\lambda 2}\\0&0&0
\end{array}\right]
$$
$$
\times
\left\{ \begin{array}{c c c}
J_0&K_0&J_0\\j'_0&K'_0&j_0\\J'&K&J
\end{array}\right\}
\left\{ \begin{array}{c c c}
\lambda'_0&K_{\lambda 0}&\lambda_0\\s&K_{s0}&s\\j'_0&K'_0&j_0
\end{array}\right\}
\left\{ \begin{array}{c c c}
J_1&K_1&J_1\\j&K'&j'\\J&K&J'
\end{array}\right\}
\left\{ \begin{array}{c c c}
\lambda'_1&K_{\lambda 1}&\lambda_1\\s&K_{s1}&s\\j'_1&K'_1&j_1
\end{array}\right\}
\left\{ \begin{array}{c c c}
\lambda'_2&K_{\lambda 2}&\lambda_2\\s&K_{s2}&s\\j'_2&K'_2&j_2
\end{array}\right\}
$$
\begin{equation}
\times
\left\{ \begin{array}{c c c}
j'_2&K'_2&j_2\\j'_1&K'_1&j_1\\j'&K'&j
\end{array}\right\} .
\end{equation}

The reduced matrix element of the electrostatic interaction in Eq.~(34) is 
defined in \cite{Jucys1972}.
In the case of  the inner-shell ionization of atoms, the state of the ion is 
not stable
and decays via radiative or Auger transition.
Then the expression (33) should be modified following the recommendations 
of Section 2.3 as it was pointed out in the Section 2.2.

The expression (33) for the electron-impact ionization can be also used for the
investigation of the ionization of atoms by protons and highly charged ions.
Then the expressions for the cross-section and reduced matrix element should
suffer some changes.
The alignment parameters for $L_3$-subshell of Cd and Sb atoms were obtained by
observing the degree of polarization of the $L_1$-lines excited by proton impact
\cite{Gelfort}.

\vspace*{ 5mm}
\noindent
{\bf 3.3. Radiative recombination}
\vspace{3mm}

The process of the radiative recombination of polarized ions with polarized electrons
can be written as follows:
\begin{equation}
A^{n+}(\alpha_0 J_0M_0)+e^-({\rm\bf p}_0 m_0)\to
A^{(n-1)+}(\alpha_1 J_1M_1) + h\nu(\epsilon_q{\rm\bf k}_0).
\end{equation}
 
From the relation of detailed balance, it follows that the cross-section of the
radiative recombination $\sigma^{rr}_{f\to i}(E_0)$ is related to 
 the cross-section of the photoionization  $\sigma^{ph}_{i\to f}(E_1)$
through detailed balance (Milne relation) as
\begin{equation}
\sigma^{rr}_{f\to i}(E_0)=\frac{(\alpha E_1)^2}{2  E_0}
\frac{g_i}{g_f}\; \sigma^{ph}_{i\to f}(E_1).
\end{equation}

\noindent
Here $E_0$ and $E_1$ are the energy of an electron  and emitted photon, respectively,
$g_i$ and $g_f$ are the statistical weights of the initial and final states, 
$\alpha$ is the
fine structure constant and $E_1=E_0+I_p$ ($I_p$ is the ionization energy).
Then the general expression for the differential radiative recombination 
cross-section can be written in the form by using Eq.~(18):
$$
\frac{d\sigma(J_0M_0{\rm\bf p}_1m_s
\to J_1M_1 \hat{\epsilon}_{q1} {\rm\bf k}_{01})}{d\Omega}
$$
$$
=
C
\sqrt{4\pi} \sum_{K_0,K_r,K_1,K_s,K_j,K,k_1,k'_1}
B^{ph}(K_0,K_1,K_r,K_\lambda,K_s,K_j,K,k_1,k'_1)
$$
$$
\times
\sum_{N_0,N_r,N_1,N_s,N_j,N}
\left[\begin{array}{ccc}
K_0 & K_j & K\\ N_0 & N_j & N
\end{array}\right]
\left[\begin{array}{ccc}
K_\lambda & K_s & K_j\\ N_\lambda & N_s & N_j
\end{array}\right]
\left[\begin{array}{ccc}
K_1 & K_r & K\\ N_1 & N_r & N
\end{array}\right]
Y^*_{K_\lambda,N_\lambda}(\hat{\rm\bf p}_1)\;
T^{*K_0}_{N_0}(J_0,J_0,M_0|\hat{J_0})
$$
\begin{equation}
\times
T^{K_r}_{N_r}(k_1,k'_1,q_1|\hat{{\rm\bf k}}_{01})\;
T^{*K_s}_{N_s}(s,s,m_s|\hat{s})\;
T^{K_1}_{N_1 N'_1}(J_1,J_1,M_1,M'_1|\hat{J_1}).
\end{equation}
\noindent
Here $C=(\pi\alpha^2 E_1^2)/E_0$.

If a recombined ion arises in excited state, it can suffer the radiative decay 
to a lower
excited or ground state by emitting a photon.
The polarization state of this photon depends on the states of the ion and electron.
For the description of the polarization characteristics of the second photon, the 
expression for the radiative recombination cross-section as that of the first-step
process is necessary.
It may be obtained following the recommendations of Section 2.3.

\vspace{ 5mm}
\noindent
{\bf  4. Decay of excited atoms}
\vspace{ 3mm}

In the case of the registration of the decay products of  formed ion,
much information can be gained  not only  about the structure of the system
under investigation but also about the process itself and
many body interactions.
Here the Auger decay process is often used \cite{Berezhko}.
But in the case of the creation of a vacancy in the outermost closed shell
of atoms containing one valence electron, the  radiative transition is only
 way of its decay.
The fluorescence and 
Auger decay of an excited ion or atom are the processes of the second step
following  inner-shell excitation or ionization of the atom.
In the present section, the expressions for the radiative and Auger decay as the
second step process will be presented.
These expressions were obtained by applying atomic theory methods 
\cite{Jucys1972,Rudzikas} and 
the graphical technique of the angular momentum \cite{Jucys1965}.

\vspace*{ 5mm}
\noindent
{\bf 4.1. Radiative decay}
\vspace{3mm}

The radiative decay of an atom or ion is the inverse process to the photoexcitation and
can be written in the form:
\begin{equation}
A(\alpha_1J_1M_1) \to
A(\alpha_2J_2M_2) + h\nu(\hat{\epsilon}_{q2}{\bf k}_{02}) .
\end{equation}

\noindent
The general expression for the probability of the fluorescence 
following the photoionization of  polarized atoms
was obtained by Kupliauskien\.{e} and Tutlys \cite{kt2004}.
In the case of fluorescence as the second step process,
the initial state $M_1$ is not registered. 
Then, the expression for the radiative transition probability  is as follows:
$$
\frac{dW^r_{K_1N_1}(\alpha_1J_1 \to
\alpha_2J_2M_2 \hat{\epsilon}_{q2}{\rm\bf k}_{02}) } {d\Omega_2} =C
\sum_{K'_r,K_2,k_2,k'_2}
{\cal A}(K_1,K'_r,K_2,k_2,k'_2)
$$
\begin{equation}
\times
\sum_{N'_r,N_2}
\left[\begin{array}{ccc}
K_1 & K'_r & K_2\\N_1 &N'_r &N_2
\end{array}\right]
T^{K_2}_{N_2}(J_2,J_2,M_2|\hat{J}_2)\;
T^{K'_r}_{N'_r}(k_2,k'_2,q_2|\hat{{\rm\bf k}}_{02}),
\end{equation}

$$
{\cal A}(K_1,K'_r,K_2,k_1,k'_2) =
( \alpha_2J_2 ||Q^{(k_2)}||\alpha_1J_1) ( \alpha_2J_2 ||Q^{(k'_2)}||\alpha_1J_1)^*
$$
\begin{equation}
\times
\left[\frac{(2K_1+1)(2J_2+1)(2k_2+1)}{2K_2+1}\right]^{1/2}
\left\{\begin{array}{ccc}
J_1 & K_1 &J_1 \\ k_2 & K'_r & k'_2 \\ J_2 & K_2 & J_2
\end{array}\right\} .
\end{equation}

\noindent
The reduced matrix element in(40) is equal to
\begin{equation}
( \alpha_2J_2 ||Q^{(k_2)}||\alpha_1J_1)= (2J_2+1)^{1/2} k_{02}
\langle\alpha_2J_2 ||Q^{(k_2)}||\alpha_1J_1\rangle
\end{equation}

\noindent 
and is defined by Eq.~(4).
The constant $C=1/(2\pi)$.

\vspace*{ 5mm}
\noindent
{\bf 4.2. Auger decay}
\vspace{3mm}

Usually Auger decay  is the second step process
\begin{equation}
A^+(\alpha_1J_1M_1) \to
A^{2+}(\alpha_2J_2M_2) +{\rm e^-}({\bf p}_2,m'_s)
\end{equation}
\noindent
following inner-shell excitation or ionization of atoms and ions.
The general expression for the probability of Eq.~(42) was obtained 
by Kupliauskien\.{e}
and Tutlys \cite{kt2003a,KT2003} in the case when the Auger transition
follows the photoionization of an atom.

When the state of the intermediate ion is not registered, the Auger transition
probability acquires the form \cite{KT2003}:
$$
\frac{dW_{K_1N_1}(\alpha_1J_1\to \alpha_2 J_2M_2{\bf p}_2m'_s)}
{d\Omega_2} =
\sum_{K',K_2,K'_\lambda,K'_s}
{\cal A}^A (K_1,K_2,K'_\lambda,K'_s,K')
\sum_{N',N_2,N'_\lambda,N'_s}
\left[\begin{array}{ccc}
K'_\lambda & K'_s & K'\\N'_\lambda &N'_s &N'
\end{array} \right]
$$
\begin{equation}
\times
\left[\begin{array}{ccc}
K_2 & K' & K_1\\N_2 &N' &N_1
\end{array}\right]
T^{K_2}_{N_2}(J_2,J_2,M_2|\hat{J}_2)\;
T^{K'_s}_{N'_s}(s,s,m'_s|\hat{s}) \;
\sqrt{4\pi} Y_{K'_\lambda N'_\lambda}(\theta_2,\phi_2)
\end{equation}

\noindent
where
$$
{\cal A}^A (K_1,K_2,K'_\lambda,K'_s,K')=
2\pi \sum_{\lambda_1,j_1,\lambda_2,j_2}
\langle \alpha_2 J_2\varepsilon_2\lambda_1(j_1)J_1||H||\alpha_1J_1\rangle
\langle \alpha_2 J_2\varepsilon_2\lambda_2(j_2)J_1||H||\alpha_1J_1\rangle^*
$$
\begin{equation}
\times
 (2J_1+1)
\left[(2\lambda_1+1) (2\lambda_2+1)
(2j_1+1)(2j_2+1)(2J_2+1)(2s+1)(2K'+1)\right]^{1/2}
$$
$$
\times
         \left\{
\begin{array}{ccc}
J_2 &j_1 & J_1\\ J_2 & j_2 & J_1\\ K_2 & K' & K_1
\end{array}
          \right\}
         \left\{
\begin{array}{ccc}
\lambda_2 &s & j_2\\ K_\lambda & K'_s & K'\\ \lambda_1 & s & j_1
\end{array} \right\}
(-1)^{\lambda_2}
\left[\begin{array}{ccc}
\lambda_1 &\lambda_2 &K'_{\lambda}\\ 0 & 0 & 0
\end{array}\right].
\end{equation}

\vspace*{ 5mm}
\noindent
{\bf 5. Dielectronic recombination}
\vspace{3mm}

\vspace{ 5mm}
The process of  dielectronic recombination (DR) can be written as follows
\begin{equation}
A^+(\alpha_0 J_0M_0)+e^-({\rm\bf p}_0 m_0) \to
A^{**}(\alpha_1 J_1)
\Bigl\{
\begin{array}{c} \to A(\alpha_2 J_2M_2) + h\nu(\epsilon_{q_2},{\rm\bf k}_{02}),\\
\to A^+(\alpha_3 J_3M_3) + e^-({\rm\bf p}_1m_1).
\end{array}
\end{equation}

\noindent
It is an example of a two-step process.
The first step is resonant electron capture. The next step is radiative or Auger decay
that were described in Section 4.
Two-step approximation for DR may be applied if the interference
 with the radiative recombination
is neglected and the summation over intermediate states $J_1M_1$ that usually occurs 
in second-order perturbation theory, is limited to a single resonance.
Then, only a summation over the magnetic substates that are not registered
is retained.
DR process is finished when the photon is emitted. 

In two-step approximation (see Section 2), the cross-section for  
DR may be written as:

$$
\frac{d\sigma(\alpha_0 J_0M_0{\rm\bf p}_0 m_0 \to \alpha_1 J_1 \to
\alpha_2 J_2M_2 \epsilon_{q_2}{\rm\bf k}_{02})}{d\Omega}
$$
$$
= 2\pi
\sum_{M_1,M'_1}
\langle \alpha_2 J_2M_2 \epsilon_{q_2}{\rm\bf k}_{02}|H'|
\alpha_1 J_1M_1\rangle
\langle \alpha_2 J_2M_2 \epsilon_{q_2},{\rm\bf k}_{02}|H'|
\alpha_1 J_1M'_1\rangle^*
$$
$$
\times
\langle \alpha_1 J_1M_1|H^e|(\alpha_0 J_0M_0{\rm\bf p}_0 m_0\rangle
\langle \alpha_1 J_1M'_1|H^e|(\alpha_0 J_0M_0{\rm\bf p}_0 m_0\rangle^*
$$
\begin{equation}
\times
[(E-E_1)^2 + \Gamma^2/4]^{-1}.
\end{equation}

\noindent
Here $H'$ and $H^e$ is the radiative decay and electrostatic interaction
operators, respectively, 
$d\Omega$ is the solid angle of the emission of radiation,
$E_1$ and $E$ is the energy of the intermediate and initial state of the system
atom+electron, respectively, and
$\Gamma$ denotes the decay width of the intermediate state that includes
both radiative and nonradiative decay channels.

In two step approximation ($E\approx E_1$), the general expression for DR (46)
in the case of the interaction of a polarized ion with a polarized
electron may be obtained by applying the methods described in \cite{krt2001,KT2003}
and is as follows:

$$
\frac{d\sigma(\alpha_0 J_0M_0{\rm\bf p}_0 m_0 \to \alpha_1 J_1 \to
\alpha_2 J_2M_2 \epsilon_{q_2},{\rm\bf k}_{02})}{d\Omega}= 
\frac{2\pi}{p_0^2}
\sum_{K_1,N_1}
W^c_{K_1,N_1}(\alpha_0 J_0M_0{\rm\bf p}_0 m_0 \to  \alpha_1 J_1)
$$
\begin{equation}
\times
\frac{d\sigma^r_{K_1,N_1}(\alpha_1 J_1 \to 
\alpha_2 J_2M_2 \epsilon_{q_2},{\rm\bf k}_{02})}{d\Omega}
[(E-E_1)^2+\Gamma^2/4]^{-1}.
\end{equation}

The resonant electron capture cross-section $W^c$ is reversed to that of Auger
decay and is defined by Eq.~(43).
The expression for the radiative decay probability $dW^r/d\Omega$ is 
the same as in Eq.~ (39).

The DR process is similar to that of resonant electron transfer and excitation (RTE). 
When highly charged ions interact with low-Z atoms, the differential cross-section
for RTE may be determined by using the cross-section of DR and momentum
approximation for the distribution of electron charge in the atom \cite{Hann,KF1991}.
Angular distribution of radiation emitted after RTE in collisions of nonpolarized
U$^{90+}$ with a graphite target was investigated in \cite{Gail}.
The process of DR with emission of two and more photons was also treated
both in nonrelativistic \cite{Balashov1994} and fully relativistic 
\cite{Zakowicz} approximations.

\vspace*{10mm}
\noindent
{\bf 6. Practical applications}
\vspace{ 3mm}

General expression for the differential cross section and probability of the
excitation and ionization of polarized atoms and ions by polarized electrons or
radiation can be used to obtain many more simple expressions applicable for 
specific experimental conditions.
The authors who used the density matrix formalism to derive the expressions for
various parameters describing the polarization usually started from 
the very beginning
by formulating the problem dealing the specific experiment.
Below a limited number of the most important papers are reviewed.

In the case of the photoionization of polarized atoms by polarized radiation, Jacobs
\cite{Jacobs1972} obtained the general expression for the differential cross-section,
the asymmetry parameters $\beta$ of the angular distribution and $\gamma$,
$\delta$ and $\xi$ for the spin polarization of photoelectrons.
The statement that the photoion was not detected was assumed from the very beginning.
The angular distribution of photoelectrons from nonpolarized atoms was investigated
in \cite{Cooper,Dill}.
The investigation of the
angular distribution of photoelectrons with specific spin orientation in the case
of nonpolarized atoms by polarized dipole radiation was carried out by Cherepkov
\cite{Cherepkov1973}.
General expression for the angular distribution and polarization of photoelectrons
from nonpolarized atoms in the region of autoionizing states was obtained in 
\cite{Kab1976}.
Here the photons of any multipolarity were treated.
Later on, the expressions for the cross-sections and asymmetry parameters for the
 angular distribution of photoelectrons from polarized atoms exposed to polarized
radiation in both resonant \cite{Baier1994} and nonresonant \cite{Klar1980,Cherepkov1995}
case were obtained.
These expressions also found their applications for the investigation of magnetic
dichroism in the angular distributions of photoelectrons \cite{Plotzke}.
Recently, dramatic nondipole effects in low-energy photoionization were discovered
both theoretically and experimentally \cite{Hemmers}.

Much larger number of papers is devoted to the investigation of Auger and 
fluorescence decay of atoms following ionization by electrons and photons.
In 1972 and 1974, three papers presented calculations on angular distributions 
of ionized atom decay products.
The polarization of characteristic radiation excited by electron impact \cite{Farlane}
and angular distribution of Auger electrons following photoionization  
\cite{Flugge,Cleff1974} were calculated did not applying the density matrix formalism.
Later on, Kabachnik and coworkers \cite{Kab1976,Berezhko1978,
Berezhko-Sizov,Kabachnik1984,Balashov1997,Kab1999,Ueda},
Lohman {\em et al} \cite{Blum1986,Lohman} and Bartschat and
Grum-Grzhimailo \cite{Bartschat} obtained numerous expressions describing the
angular distribution and spin polarization of Auger electrons for various specific
experimental conditions.
As the expressions were hinted to the applications of noble gaseous the initial state
of the atom was considered randomly orientated.
The atoms could be ionized by polarized dipole photons \cite{Berezhko1978} or
nonpolarized electrons \cite{Berezhko,Berezhko-Sizov,Blum1986,Lohman}.
Angular distribution of Auger electrons ejected by electron-impact from laser-excited
and polarized atoms was described in \cite{Balashov1997}.
A resonant cascade model based on a stepwise approach was suggested 
\cite{Kab1999} for the analysis of the angular correlation in the decay
of core-excited resonances produced by photoabsoption.
The model was applied to the description of angular distributions of Auger 
electrons and fluorescence.

The alignment of atoms and ions in radiative electron capture (REC) into
ground and excited states of ions is recently formed field for the polarization
investigation in electron-ion interactions \cite{Eichler1998,Eichler2002,
Surzhykov}.
The angular distribution of the decay radiation of recombined ions may provide 
useful information about the ionic sublevels following electron capture from
atoms \cite{Fainstein}, molecules and solids.
But the parameters of the resonant electron transfer and excitation (RETE) are 
similar to those of dielectronic recombination. 
Some special cases of RETE were investigated in \cite{Gail,Tanuma}.

The general expressions presented in Sections 2--5 can be used for the description
of the processes mentioned above that were investigated with the help of the
density matrix formalism.
These general expressions are written in the invariant form of the expansions over
the state multipoles, i.e.  they are independent of the choice of coordinate system.

If the atoms or electrons in the initial state are randomly orientated, the general
expressions have to be averaged over the projections of the total angular 
momentum $J$ or electron spin, respectively.
When the final states of the products of processes are not detected, the summation over
projections over the states of the atom or electron spin should be performed.
The integration over the angles of the emission of electrons or radiation should
be performed if they are not detected.

The examination of the general expressions (11), (18), (20), (30), (33), (37), (39)
and (41) shows that only the tensor
$T^K_N(J,J,M|\hat{J})$ depends on the projection $M$. 
Summation over $M$ gives \cite{krt2001}

\begin{equation}
\sum_{M} T^{K}_{N}(J,J,M|\hat{J}) =\delta(K,0)\delta(N,0).
\end{equation}

\noindent
Integration over the angles leads to \cite{VMK1988}
\begin{equation}
\int_0^\pi \sin \theta d\theta \int_0^{2\pi} d\phi Y_{KN}(\theta\phi)=
\sqrt{4\pi}\delta(K,0)\delta(N,0).
\end{equation}

A more complicated case occurs with the polarization of radiation.
The nonpolarized radiation usually is represented as a sum of equal parts of the
left- and right-polarized radiation since the helicity $q\neq 0$.
Thus for the nonpolarized dipole photon we have:
\begin{equation}
\sum_{q=\pm1} T^{K}_{N}(1,1,q|\hat{{\rm \bf k_0}}) =
\frac{1}{3}\delta(K,0)\delta(N,0) + \sqrt{\frac{4\pi}{3}}
\left[ \begin{array}{ccc}
1 &1 & 2\\ 1 & -1 & 0
\end{array} \right ]
Y_{2N}(\theta,\phi) \delta(K,2).
\end{equation}

\noindent
This expression can also be used in the case when the polarization of the emitted radiation
is not detected.

\vspace{ 5mm}
{\bf 6.1. The angular distribution of photoelectrons from unpolarized atoms}
\vspace{3mm}

If the initial state of  atoms is randomly orientated,
the final state of  ions
and the spin of the photoelectrons are not registered, the summation over
the components of the ion state and the spin of photoelectron, and
averaging over the components of the atoms is necessary.
From the application of Eq.~(48) in (18)  follows that
$K_0=K_1=K_s=0$, $K_r=K_\lambda=K_j=K$.
The use of these values in Eq.~(18) and the choice of the z axis along the
direction of incoming radiation allows us to write the differential
cross-section
$$
\frac {d\sigma(\alpha_0J_0q \to \alpha_1J_1)}{d\Omega_p}=
\frac{\pi}{2J_0+1}
\sum_{K,k,k'} {\cal B}^{ph}(0,0,K,K,0,K,K,k,k')
(-1)^{k-q} \left[\frac{2J_1+1}{2k+1}\right]^{1/2}
\left[\begin{array}{ccc} k&k'&K\\q& -q& 0 \end{array}\right]
$$
\begin{equation}
\times
P_K(\cos \theta).
\end{equation}
\noindent
In Eq.~(51), $P_K(\cos\theta)$ stands for a Legendre polynomial.
For the  circular polarization of incoming dipole radiation ($k=k'=q=1$),
the cross-section may be written:

\begin{equation}
\frac {d\sigma(\alpha_0J_0 \to \alpha_1J_1)}{d\omega_p}=
\frac{\sigma(J_0\to J_1)}{4\pi}\left[1-\frac{1}{2}
\beta P_2(\cos \theta) \right]
\end{equation}
\noindent
where
\begin{equation}
\beta =\frac{1}{\sqrt{2}} \frac{{\cal B}^{ph}(0,0,2,2,0,2,2,1,1)}
{{\cal B}^{ph}(0,0,0,0,0,0,1,1)}, 
\end{equation}
\begin{equation}
\sigma(\alpha_0J_0 \to \alpha_1J_1)=\frac{4\pi}{3}
\frac{[2J_1+1]^{1/2}}{2J_0+1} 
{{\cal B}^{ph}(0,0,0,0,0,0,1,1)}.
\end{equation}

The expressions for $\beta$ (53) and $\sigma$ (54) differ from those
obtained in \cite{krt2001} and used in the calculations of the
angular distribution of 2p photoelectrons from Na atoms in the ground
\cite{Kupl2001} and excited states \cite{RK1998,Kupl2001,KL2001}.
The spin polarization parameters $\gamma$, $\delta$ and $\xi$ were also
calculated for the 2p photoionization of Na in the ground and first
excited states \cite{JK2001}.

\vspace{ 5mm}
{\bf 6.2. Angular distribution of Auger electrons for unpolarized atoms}

\vspace{ 3mm}
The process of the photoionization of an atom $A$
with following Auger decay of a photoion $A^+$  in a two-step approximation can
be written as follows:

\begin{eqnarray}
A(\alpha_0J_0M_0) + h\nu(\hat{\epsilon}_q,{\bf k}_{01}) & \to &
A^+(\alpha_1J_1M_1) + {\rm e^-}({\bf p_1},m_s) \nonumber\\
& \to &
A^{2+}(\alpha_2J_2M_2) + {\rm e^-}({\bf p}_1,m_s) +
{\rm e^-}({\bf p}_2,m'_s).
\end{eqnarray}

\noindent
In Eq.~(55), $\alpha_0$, $\alpha_1$ and
$\alpha_2$ indicate the configuration and other quantum numbers,
$J_0M_0$, $J_1M_1$ and $J_2M_2$
describe the total angular momenta and their magnetic components
of the electron cloud of an atom in the initial state, intermediate
photoion and doubly charged ion in the final state, respectively.
The photo- and Auger electrons have the momentum
${\bf p}_1$ and ${\bf p}_2$, the projection of their spin $s$ is $m_s$
and $m'_s$, respectively.
The wave vector of incoming radiation is indicated by ${\bf k}_{01}$
($|{\bf k}_{01}|=\omega_1/c$,
where $\omega_1$ is the frequency of radiation.

An  expression for the cross-section of the processes (55)  was
derived by Kupliauskien\.{e} and Tutlys
\cite{KT2003,kt2003a} and can be written in the form:

$$
\frac{dW(\alpha_0J_0M_0 \hat{\epsilon}_q {\rm\bf k}_{01}\to
 \alpha_1J_1{\bf p}_1m_s \to
\alpha_2J_2M_2{\bf p}_2m'_s)} {d\Omega_1 d\Omega_2}
$$
\begin{equation}
=
\sum_{K_1,N_1}
\frac{d\sigma_{K_1N_1}(\alpha_0J_0M_0  \hat{\epsilon}_q {\rm\bf k}_{01} \to 
\alpha_1J_1{\bf p}m_s)}
{d\Omega_1}
\cdot
\frac{dW_{K_1N_1}(\alpha_1J_1 \to
\alpha_2J_2M_2{\bf p}_2m'_s)} {d\Omega_2} .
\end{equation}

\noindent
Here $d\Omega_1$ and $d\Omega_2$ indicate the solid angles of the emission of
the photoelectron and Auger electron, respectively.

The probability describing the angular distribution of Auger electrons from
unpolarized atoms simplifies as a result of
the summation  of (56) over the magnetic components of the
spins of photo- and Auger
electrons, the total angular momentum of a doubly charged ion, integration
over the angles of photoelectron and averaging over the magnetic components
of an atom.
 In the case of dipole approximation, it acquires the following form:
$$
\frac{dW(J_0\to J_1\to J_2 {\bf p}_2)}{d\Omega_2}
=\sum_{K_1} {\cal  A}^A(K_1,0,K_1,0,K_1) \;
B(K_1)\; P_{K_1}(\cos\theta_2)
$$
\begin{equation}
={\cal A}^A(0,0,0,0,0)B(0)[1 + \beta P_2(\cos \theta_2)].
\end{equation}

\noindent
Here 
\begin{equation}
B(K_1)=\frac{4\pi}{2J_0+1}
\left[\frac{2K_1+1}{3}\right]^{1/2}
\left[\begin{array}{ccc} 1& 1&K_1\\1&-1&0\end{array}\right]
{\cal B}(K_1,0,K_1,0,0,0,K_1,1,1),
\end{equation}
\begin{equation}
\beta = \frac{{\cal A}^A(2,0,2,0,2) B(2)}{{\cal A}^A(0,0,0,0,0) B(0)}
= \frac{{\cal A}^A(2,0,2,0,2)}{{\cal A}^A(0,0,0,0,0)} A_2
\end{equation}
\noindent
where $A_2$ is the alignment \cite{krt2001}.
Its expression coincides with that presented by Kabachnik and Sazhina
\cite{Kabachnik1984} if the expressions for $B(K_1)$ would be inserted.

In the case of unpolarized atoms,
the expression for the probability describing angular correlations between
photo- and Auger electrons is obtained by summation of Eq.~(56) over magnetic
components of the spins of photo- and Auger electrons, the angular momentum
of a doubly charged ion and averaging over magnetic components of an atom.
For circularly polarized dipole ionizing radiation, it has the following expression:

$$
\frac{dW(J_0\to J_1{\bf p_1}\to J_2 {\rm \bf p}_2)}
{d\Omega_1 \; d\Omega_2} =
 \sum_{K_1N_1}  {\cal A}(K_1,0,K_1,0,K_1) \;
\left[\frac{4\pi}{2K_1+1}\right]^{1/2}Y_{K_1N_1}(\theta_2,\phi_2)
$$
\begin{equation}
\times
\sum_{K_r,K_\lambda} B'(K_1,K_\lambda,K_r)
\sum_{N_r,N_\lambda}
\left[ \begin{array}{ccc}
K_1 & K_\lambda & K_r \\  N_1 & N_\lambda & N_r
\end{array} \right]
\frac{4\pi}{\sqrt{2K_r+1}}
Y^*_{K_rN_r}(\theta_0,\phi_0) \;
Y_{K_\lambda N_\lambda}(\theta_1,\phi_1).
\end{equation}

\noindent
Here
$$
B(K_1,K_\lambda,K_r)=
\frac{1}{2J_0+1}\left[\frac{(2K_1+1)(2K_r+1)}3\right]^{1/2}
\left[\frac{2K_1+1}3\right]^{1/2}
\left[\begin{array}{ccc} 1& 1&K_r\\1&-1&0\end{array}\right]
$$
\begin{equation}
\times
{\cal B}(K_1,0,K_r,K_\lambda,0,K_\lambda,K_1,1,1).
\end{equation}

More examples of the practical application of the general expressions can be 
found in Refs.~\cite{krt2001,Kup2003,kt2003a,KT2003,kg2003,kt2004}.

\vspace*{ 5mm}
\noindent
{\bf 7. Concluding remarks}

\vspace{ 3mm}
A method for the derivation of general expressions for the cross-section 
and transition
probability describing the polarization states of all particles participating in the 
interaction of polarized photons and electrons with polarized atoms is developed.
It is alternative to the density matrix formalism.
The graphical technique of the angular momentum applied for the integration over
angular and spin variables of the matrix elements as well as for the expansion of
the cross-sections and probabilities over the spherical tensors enables us to obtain
the most general expressions for the cross-sections.
Irreducible tensors are selected for the description of polarization because they 
have the simplest possible behavior under changes of directions.
The method is also generalized for the multistep processes.

The following processes playing very important role in plasma are investigated:
\begin{itemize}
\item
excitation and ionization of atoms by photons;
\item
excitation and ionization of atoms by electron impact;
\item
radiative recombination of an ion and electron;
\item
radiative and Auger decay of excited and ionized atoms.
\end{itemize}

The dielectronic recombination of an ion with an electron is an example 
of two-step process.

The practical application of the general expressions for the description of
more simple processes under specific experimental conditions is easy to
accomplish.
The asymmetry parameter of the angular distribution of photoelectrons and
Auger electrons following photoionization of unpolarized atoms as well as
the parameters describing the angular correlations between the photo- and
Auger electrons are obtained.
These examples demonstrate the way for the derivation of more simple 
expressions.
All expressions are presented in a similar form convenient for development 
of computer software and practical applications.

\vspace*{5mm}

\begin{center}
{\large\bf  ATOMO TEORIJOS METODAI ATOM\n{U} S\n{A}VEIKAI SU FOTONAIS
IR ELEKTRONAIS TIRTI}
\vspace{3mm}

A.Kupliauskien\.{e}
\vspace{3mm}

{\small\em VU Teorin\.{e}s fizikos ir astronomijos institutas, Vilnius, Lietuva}
\end{center}
\vspace{3mm}

\noindent
{\bf Santrauka}

Sklaidos u\v{z}daviniuose sklaidomosios dalel\.{e}s kryptis yra apibr\.{e}\v{z}ta, 
tod\.{e}l reakcijos produkt\n{u} b\={u}sen\n{u}
u\v{z}pildai  b\={u}dinga asimetrija \v{s}ios krypties at\v{z}vilgiu.
Iki \v{s}iol tokia poliarizacija ir asimetrija buvo tiriamos
tankio matricos metodais.
Prie\v{s} ketvert\n{a} met\n{u} poliarizacijos rei\v{s}kiniams, kai atomai
ir jonai s\n{a}veikauja su elektronais ir fotonais, nagrin\.{e}ti buvo
pritaikyti atomo teorijos metodai, kurie iki \v{s}iol buvo taikomi
izoliuotiems atomams tirti, neatsi\v{z}velgiant \n{i} i\v{s}skirt\n{a}
krypt\n{i} erdv\.{e}je.
Ap\v{z}velgti darbai, skirti poliarizuot\n{u} atom\n{u}
s\n{a}veikos su poliarizuotais fotonais ir elektronais skerspj\={u}vi\n{u}
bendriausioms i\v{s}rai\v{s}koms nereliatyvistiniu art\.{e}jimu surasti.
Jud\.{e}jimo kiekio momento grafin\.{e} technika \v{s}uolio pritaikyta
\v{s}oulio operatori\n{u} matricini\n{u} element\n{u} kvadratams
integruoti kampini\n{u} ir sumuoti
sukinini\n{u} kintam\n{u}j\n{u} at\v{z}vilgiu.
Skers\-pj\={u}vi\n{u} i\v{s}rai\v{s}kos u\v{z}ra\v{s}ytos daugialypiais
sferini\n{u} multipoli\n{u} skleidiniais, kadangi sferiniai tenzoriai
transformuojasi papras\v{c}iausiai, kei\v{c}iant matavimo kryptis.
Pateiktas b\={u}das foton\n{u} ar elektron\n{u} spinduliuotei po atomo ar
jono jonizacijos, su\v{z}adinimo ar rekombinacijos dvipakopiu
art\.{e}jimu nagrin\.{e}ti.

I\v{s}nagrin\.{e}ti svarbiausi plazmoje vykstantys  vyksmai: 
atom\n{u} ir jon\n{u} su\v{z}adinimas ir jonizacija fotonais bei elektronais,
jono ir elektrono fotorekombinacija, spinduliavimo ir Auger
\v{s}uoliai su\v{z}adin\-tuose ir jonizuotuose atomuose.
Dvipakop\n{i} vyksm\n{a} iliustruoja dvielektron\.{e} rekombinacija.
Taip pat parodyta, kaip galima surasti skerspj\={u}vi\n{u} i\v{s}rai\v{s}kas
konkretiems eksperimentams apra\v{s}yti. Vis\n{u} i\v{s}rai\v{s}k\n{u} forma
paprasta, patogi kompiuterin\.{e}ms programoms ra\v{s}yti.

\end{document}